\documentclass[a4paper,11pt]{article}

\usepackage[T1]{fontenc}
\usepackage{jheppub}

\title{\boldmath From Bubble Collisions to Sound Waves: the Role of Bubble Oscillations}

\author[b,*]{Guangshang Chen,%
\note[*]{All authors contributed equally to this work.}}

\author[a,*,\dagger]{Jianpeng Dai,%
\note[$\dagger$]{Corresponding author.}}

\author[c,*,\ddagger]{Yang Xiao,%
\note[$\ddagger$]{Corresponding author.}}

\author[c,*]{Yang Zhang}

\affiliation[a]{School of Data Science and Artificial Intelligence, Wenzhou University of Technology, Wenzhou, China}

\affiliation[b]{Key Laboratory of Particle Physics and Particle Irradiation (MOE), Institute of Frontier and Interdisciplinary Science,
Shandong University, Qingdao 266237, P. R. China}

\affiliation[c]{School of Physics, Henan Normal University,
Xinxiang 453007, P. R. China}

\emailAdd{chenguangshang@sdu.edu.cn}
\emailAdd{daijianpeng@wzut.edu.cn}
\emailAdd{xiaoyangphy@gmail.com}
\emailAdd{zhangyang2025@htu.edu.cn}

\abstract{%
Sound waves generated by expanding bubbles during cosmological first-order phase transitions can persist long after bubble collisions and constitute an important source of gravitational waves (GWs). These waves are often assumed to propagate freely after bubble collisions. However, residual scalar field oscillations can continue to transfer energy and momentum to the fluid as the field relaxes toward the true vacuum. Using the harmonic approximation to describe these oscillations, we derive the resulting acoustic response and incorporate it into effective initial conditions for free acoustic propagation. We test this description using three-dimensional lattice simulations of the coupled scalar–fluid system. In the single bubble case, using these effective initial conditions near the end of the phase transition reproduces the subsequent GW spectrum with errors of a few percent relative to the full simulation. For multiple bubbles, collisions occurring at different times reduce the effectiveness of these initial conditions. These findings clarify the transition from bubble collisions to free acoustic propagation, contributing to a more complete theoretical understanding of phase transition dynamics.
}

\begin{document}
\maketitle
\raggedbottom

\section{Introduction}
\label{sec:introduction}

Cosmological phase transitions provide an important connection between the thermal history of the early Universe and physics beyond the Standard Model \cite{Athron2024,LiGong:2024qmt,Mazumdar:2018dfl}. The electroweak phase transition is of particular interest, as new physics can turn the smooth crossover predicted by the Standard Model into a first-order phase transition, providing the out-of-equilibrium conditions required for electroweak baryogenesis \cite{DOnofrio2014,Morrissey2012,White:2016nbo,Trodden:1998ym}. Such a transition proceeds through the nucleation, expansion, and collision of bubbles, transferring part of the released vacuum energy to the surrounding plasma and generating bulk fluid motion \cite{Espinosa2010,Cai:2020djd,Wang:2022lyd}. The resulting anisotropic stresses can generate a stochastic gravitational wave (GW) background, offering a possible way to probe electroweak-scale new physics with future space-based observatories such as LISA, Taiji, and TianQin \cite{amaroseoane2017laserinterferometerspaceantenna,Ruan2020,Luo2016}. To achieve this goal, an accurate theoretical understanding of the GW production mechanisms and the underlying phase transition dynamics is essential~\cite{Gowling2023,Caprini2016,Caprini2020}. Among the GW sources associated with cosmological first-order phase transitions, sound waves induced by expanding bubbles are particularly important, as they can persist long after bubble collisions and provide a major contribution to the GW signal \cite{Hindmarsh2014,Hindmarsh2015,Ellis2020}.

The role of acoustic motion in GW production from first-order phase transitions was recognized in early studies \cite{Hogan1986}. Subsequent three-dimensional simulations of coupled scalar-fluid dynamics demonstrated that the fluid motion induced by expanding bubbles persists after bubble collisions and continues to source GWs \cite{Hindmarsh2014,Hindmarsh2015,HindmarshShape2017,Correia2025}. Such simulations have also been extended to strong transitions in
an expanding Universe~\cite{Zou:2026hnb}. Building on previous numerical results, the sound shell model (SSM) provided an analytic description of acoustic GW production by treating the acoustic field as a superposition of freely propagating sound shells, with their initial profiles determined by the self-similar fluid motion induced by steadily expanding bubbles \cite{HindmarshSSM2018,HindmarshHijazi2019,RoperPol2024}. To avoid the numerical challenge of resolving the widely separated characteristic scales of the scalar field and fluid motion, hybrid simulations evolved one-dimensional, spherically symmetric velocity and enthalpy profiles after bubble collisions and combined them through linear superposition to construct three-dimensional acoustic fields \cite{JinnoHybrid2021}. This strategy was further developed in Higgsless simulations, which directly solve the three-dimensional fluid equations while representing expanding bubbles as prescribed sources of energy and momentum \cite{JinnoHiggsless2023}. Complementing these numerical approaches, the hydrodynamic SSM extended the original model by accounting for the driving of sound shells during their initial collisions, when parts of the expanding bubble interfaces remain uncollided and still drive the surrounding fluid \cite{Cai2023}. In parallel, studies of the SSM have examined the effects of bubble nucleation and lifetime distributions~\cite{Guo:2023gwv} and developed
improved spectral fits~\cite{Guo:2024gmu}.
Extensions to expanding backgrounds have clarified the role of the
source lifetime and expansion history~\cite{Guo:2020grp,Xiao:2024rsj},
while fluid dissipation can further modify the acoustic lifetime
and the resulting GW spectrum~\cite{Guo:2023koq,Guo:2026ofc}. More recently, the initial velocity spectrum for free acoustic propagation has been refined by following fluid perturbations during bubble expansion and using kinetic energy conservation to determine an effective onset of the acoustic stage \cite{Caprini2026}. Together, these developments have improved the description of sound wave production and evolution during bubble expansion and collision.

In many of these approaches, sound waves are assumed to propagate freely after bubble collisions. However, the completion of bubble collisions does not necessarily imply that the scalar field no longer affects the subsequent fluid evolution. Numerical studies of vacuum bubble collisions have shown that the scalar field can retain kinetic and gradient energy and continue to oscillate around the true vacuum \cite{Cutting2018,Gould2021}. In the presence of a plasma, damping of these oscillations can transfer additional energy and momentum to the fluid \cite{Hindmarsh2015,Ekstedt2026}. As a result, previously generated sound waves can remain driven by the residual scalar field after collisions. The timing and duration of this additional driving can further modify their evolution. Neglecting the scalar source immediately after collisions may therefore lead to inaccurate initial conditions for free acoustic propagation and, consequently, affect predictions of the resulting GW spectrum. This raises the question of how residual scalar field oscillations drive the existing sound waves and whether their effects can be captured by an effective description.

In this work, we investigate how residual scalar field oscillations after bubble collisions affect the subsequent acoustic evolution. Starting from energy-momentum conservation, we develop a theoretical model to describe the fluid response to the remaining scalar oscillations. This model provides the basis for an effective description that incorporates the effects of residual driving into the initial conditions for free acoustic propagation. We test this approach against three-dimensional collision simulations and quantify the errors introduced by the effective description. The remainder of this paper is organized as follows. Sec.~\ref{sec:mismatch} derives the acoustic equations and examines the fluid response to residual scalar field driving. Sec.~\ref{sec:initialcondition} constructs effective initial conditions for subsequent free acoustic propagation. Sec.~\ref{sec:numeric} compares this effective description with three-dimensional collision simulations and quantifies how the matching error and the remaining scalar contribution depend on the matching time. Sec.~\ref{sec:conclusions} summarizes our conclusions. 


\section{Scalar field dynamics and acoustic response}
\label{sec:mismatch}

\subsection{Scalar-fluid dynamics in the weak-fluid limit}

We model the first-order phase transition using a scalar order parameter $\phi$ coupled to a relativistic perfect fluid representing the thermal plasma~\cite{Ignatius:1993qn,Hindmarsh2015}. Since the timescales of interest are much shorter than the Hubble time, we neglect cosmological expansion and work in Minkowski spacetime with the metric $g_{\mu\nu}=\operatorname{diag}(-1,1,1,1)$.

The total energy-momentum tensor of the system can be decomposed into scalar field and fluid contributions, $T_{\mu\nu}=T^\phi_{\mu\nu}+T^f_{\mu\nu}$, with
\begin{align}
\label{eq:emt_phi}
T^\phi_{\mu\nu} &= \partial_\mu\phi\,\partial_\nu\phi-g_{\mu\nu}\left[\frac12(\partial\phi)^2+U(\phi)\right],\\
\label{eq:emt_fluid}
T^f_{\mu\nu} &= wu_\mu u_\nu+pg_{\mu\nu},
\end{align}
where $w=e+p$ is the enthalpy density, $e$ and $p$ are the fluid energy density and pressure, respectively, and $u^\mu=\gamma(1,\mathbf u)$ is the fluid four-velocity with the Lorentz factor $\gamma=(1-|\mathbf u|^2)^{-1/2}$. The scalar potential $U(\phi)$ is defined as
\begin{equation}
U(\phi)=V(\phi,T_*)-V(\varphi_v,T_*),
\label{eq:fixed-potential}
\end{equation}
where $V(\phi,T)$ is the finite-temperature effective potential and $\varphi_v$ is the true vacuum field value at $T_*$. For sufficiently weak phase transitions, temperature variations between the two phases and during the subsequent relaxation are typically small. We therefore approximate the local temperature dependence of the scalar potential by evaluating it at a characteristic temperature $T_*$, which may be chosen as the nucleation or percolation temperature. Under this approximation, the free-energy density takes the separable form $\mathcal F(T,\phi)=\mathcal F_{\rm pl}(T)+U(\phi)$, and the fluid is described by a barotropic equation of state $p=p(e)$.


Following Ref.~\cite{Hindmarsh2015,Kurki-Suonio:1995yaf}, we introduce a phenomenological dissipative coupling between the scalar field and the fluid, so that the equation of motion for the scalar field takes the form
\begin{equation}
\Box\phi-U'(\phi)=\eta u^\mu\partial_\mu\phi,
\label{eq:scalar-eom}
\end{equation}
where $\eta>0$ is a constant damping coefficient. Combined with total energy-momentum conservation, Eq.~\eqref{eq:scalar-eom} determines the fluid dynamics through
\begin{equation}
\partial^\mu T^f_{\mu\nu}= -\eta u^\mu\partial_\mu\phi\,\partial_\nu\phi.
\label{eq:exchange-current}
\end{equation}
Consequently, the same interaction that damps the scalar field thus transfers energy and momentum to the plasma. The amplitude of the resulting fluid response is controlled by the characteristic scalar energy scale relative to the background plasma enthalpy, motivating the weak fluid approximation developed below.

In the weak fluid regime, the background plasma enthalpy $w_0=e_0+p_0$ is large compared with the characteristic energy density of the scalar. Take its inverse as a power-counting parameter, the scalar and fluid variables admit the large-enthalpy expanded as
\begin{equation}\label{eq:large-enthalpy-expansion}
    \begin{aligned}
    \mathbf u &= \frac{\mathbf v}{w_0}+\frac{\mathbf v_1}{w_0^2}+O(w_0^{-3}), \\
    e &= e_0+d+\frac{d_1}{w_0}+O(w_0^{-2}), \\
    p &= p_0+c_s^2d+\frac{1}{w_0}\left(c_s^2d_1+\frac{\kappa}{2}d^2\right)+O(w_0^{-2}), \\
    \phi &= \varphi_v + \chi +\frac{\psi}{w_0}+O(w_0^{-2}), 
    \end{aligned}
\end{equation}
where $c_s^2=p'(e_0)$ is the squared sound speed and $\kappa=w_0p''(e_0)$ characterizes the leading nonlinear correction to the equation of state. The dimensionless derivatives of the equation of state are held fixed in the expansion. Here $\varphi_v$ represent the true vacuum and $\chi$ describes the leading nonlinear scalar evolution. Four vectors $(d,\mathbf v)$ and $(d_1,\mathbf v_1)$ are the leading and next-to-leading fluid perturbations, respectively, and $\psi$ represents the scalar correction induced by fluid back-reaction.


Substituting the expansions into Eqs.~\eqref{eq:scalar-eom} and \eqref{eq:exchange-current}, we obtain the leading-order equations of motion for the scalar field and the fluid,
\begin{align}
\ddot\chi-\partial^i\partial_i\chi+U'(\varphi_v+\chi) &= -\eta\dot\chi,
\label{eq:leading-scalar}\\
\dot d+\partial_iv^i &= \eta\dot\chi^2,
\label{eq:leading-fluid-energy}\\
\dot{v}_i+c_s^2\partial_i d &= -\eta\dot\chi\partial_i\chi.
\label{eq:leading-fluid-momentum}
\end{align}
At this order, the scalar field retains its nonlinear self-interactions but evolves independently of the fluid perturbations. The fluid response is governed by linear acoustic equations with source terms determined by the scalar field evolution. From Eq.~\eqref{eq:leading-scalar}, we also obtain the evolution equations for the scalar field energy and momentum densities,
\begin{align}
\dot E_\phi+\partial_ig^i_\phi &= -\eta\dot\chi^2,
\label{eq:leading-scalar-energy}\\
\dot g_{\phi,i}+\partial^jT^\phi_{ij} &= \eta\dot\chi\,\partial_i\chi,
\label{eq:leading-scalar-momentum}
\end{align}
where $E_\phi$, $g_{\phi,i}$, and $T^\phi_{ij}$ are the scalar field energy density, momentum density, and spatial stress, respectively, evaluated at leading order.


Beyond leading order, the large-enthalpy expansion incorporates the back-reaction of the induced fluid motion on the scalar field. The first scalar correction satisfies
\begin{equation}
\left[\partial_t^2-\nabla^2+\eta\partial_t+U''(\varphi_v+\chi)\right]\psi=-\eta v_i\partial^i\chi.
\label{eq:scalar-feedback}
\end{equation}
Together with nonlinear fluid transport, it triggers the next-to-leading contributions of fluid response. In the laboratory-frame, the correction of energy and momentum densities is denoted as
\begin{equation}
    E_1=d_1+|\mathbf v|^2, \qquad \mathbf W=\mathbf v_1+(1+c_s^2)d\mathbf v,
\end{equation}
which are determined by the following equations:
\begin{equation}
    \begin{aligned}
        \dot E_1+\partial_iW^i &= \eta\dot\chi(2\dot\psi + v^i\partial_i\chi), \\
        \dot{W}_i+\partial_j(v_iv^j)+\partial_i\left(c_s^2d_1+\frac{\kappa}{2}d^2\right) &= -\eta\left[(\dot\psi+v_j\partial^j\chi)\partial_i\chi+\dot\chi\partial_i\psi\right].
    \end{aligned}
\end{equation}
Notably, the scalar backreaction contributes to the total anisotropic stress at the same inverse-enthalpy order as the leading quadratic fluid stress. Both contributions are therefore relevant to a consistent expansion of the total GW source.


\subsection{Stress mismatch between scalar and fluid}
\begin{figure}[t]
    \centering
    \includegraphics[width=0.9\textwidth]{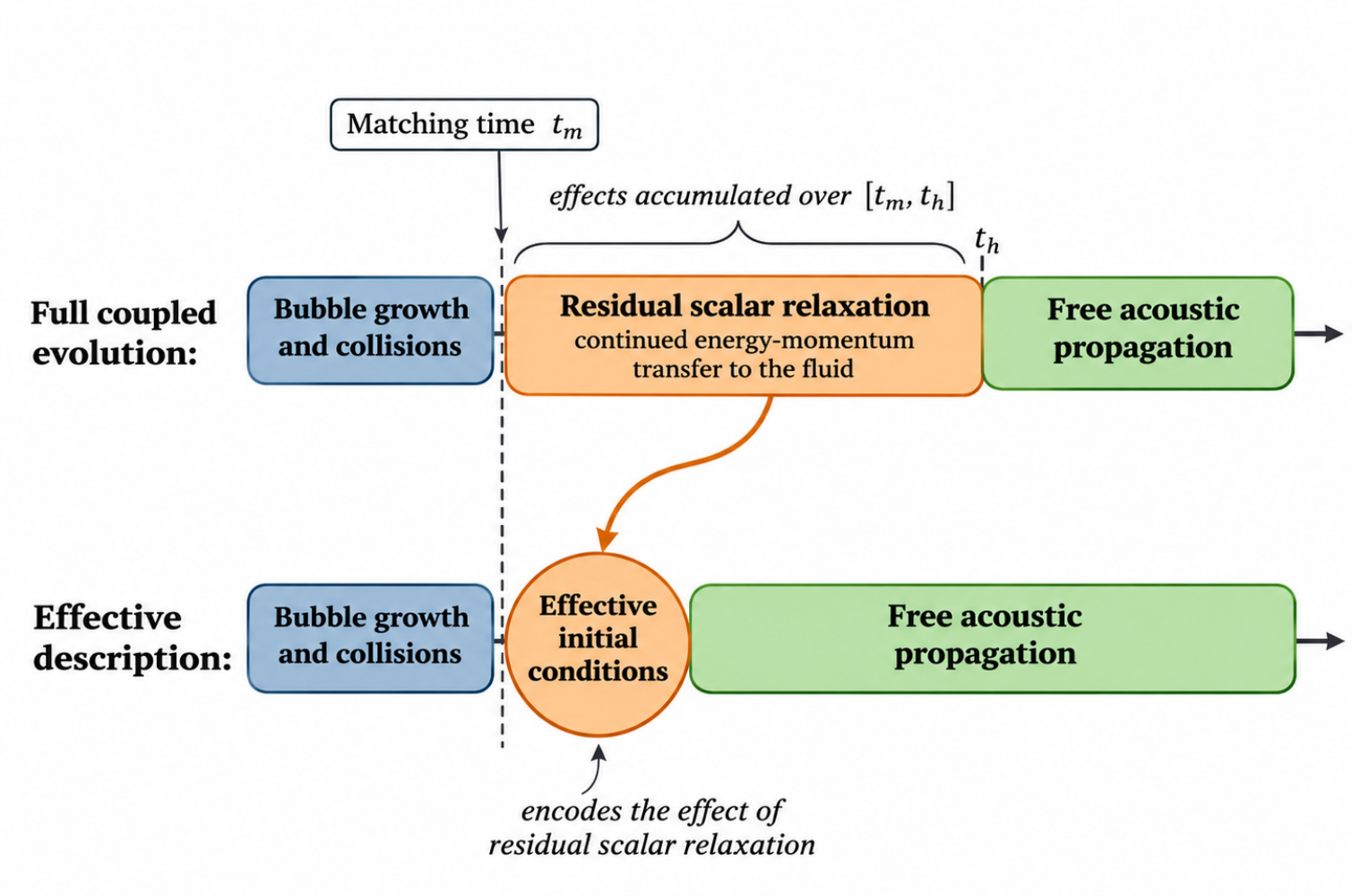}
    \caption{Schematic illustration of the effective description, in which the effects of residual scalar field relaxation over $[t_m,t_h]$ are incorporated into the initial conditions for free acoustic propagation at $t_m$.}
    \label{fig:matching_schematic}
\end{figure}

In many models of acoustic GW production, sound waves are assumed to propagate freely after bubble collisions. The equations derived above, however, describe how the residual scalar field continues to affect the fluid through energy and momentum exchange during its relaxation. Neglecting this process may therefore introduce deviations in the predicted GW spectrum. Our aim is thus to incorporate the effects of scalar field relaxation over the interval $[t_m,t_h]$ into effective initial conditions for free acoustic propagation at $t_m$, hereafter referred to as the matching time, as illustrated in Fig.~\ref{fig:matching_schematic}. We first consider the limiting case of instantaneous conversion, corresponding formally to $t_h=t_m$, in which all the remaining energy and momentum of the scalar field are transferred to the fluid at $t_m$. As we show below, although this prescription preserves the total energy and momentum densities, it generally introduces a mismatch in the spatial stress and therefore fails to reproduce the correct acoustic evolution.

To quantify the discrepancy, we first consider the total energy-momentum tensor at the leading order expressed as
\begin{equation}
    T^{\rm LO}_{ij} = (p_0 + c_s^2 d)\delta_{ij} + T^{\phi}_{ij}.
\end{equation}
The scalar contribution depends on the kinetic and potential energies of the field
\begin{equation}
    T^{\phi}_{ij} = \partial_i\chi\,\partial_j\chi + \delta_{ij}\left[\frac12\dot\chi^2-\frac12|\nabla\chi|^2-U(\varphi_v+\chi)\right].
\end{equation}
At the matching time, all the remaining scalar energy and momentum are instantaneously transferred to the fluid via a hypothetical conversion, such that 
\begin{equation}\label{eq:instantaneous-conversion}
\begin{aligned}
    \bar d(t_m)&=d(t_m)+E_\phi(t_m), \\ 
    \bar v^i(t_m)&=v^i(t_m)+g_\phi^i(t_m).
\end{aligned}
\end{equation}
The converted fluid $(\bar d, \bar{\mathbf{v}})$ then carries the spatial tensor
\begin{equation}
    \bar T_{ij} = \left[p_0 + c_s^2(d+E_\phi)\right]\delta_{ij}.
\end{equation}
Although this prescription preserves the total energy and momentum densities $T_{0\mu}$ of the system, it generally causes stress mismatch
\begin{equation}\label{eq:stress-mismatch}
    \Delta_{ij} = \bar T_{ij} - T^{\rm LO}_{ij} = c_s^2 E_\phi\delta_{ij} - T^\phi_{ij}.
\end{equation}

With the leading-order scalar energy density and spatial stress, we can explicitly separate the contributions of kinetic, gradient, and potential energies as
\begin{equation}
\Delta_{ij}=\delta^{ij}\left[\frac{c_s^2-1}{2}\dot\chi^2+\frac{1+c_s^2}{2}|\nabla\chi|^2+(1+c_s^2)U(\varphi_v+\chi)\right]-\partial_i\chi\,\partial_j\chi.
\label{eq:stress-mismatch-explicit}
\end{equation}
The kinetic and potential energies contribute differently to the scalar stress and the acoustic pressure, while spatial gradients introduce directional stresses absent from the leading-order fluid response. The mismatch therefore reflects the distinct stress--energy relations of the two components, rather than merely the anisotropy of the scalar stress. It can remain nonzero even for spatially homogeneous scalar motion, although only its spatial divergence affects the momentum evolution.

On the other hand, $\Delta_{ij}$ above can also be defined along the original coupled evolution, allowing us to examine its dynamical consequences beyond the matching surface. At leading order, total energy–momentum conservation gives
\begin{align}
    \partial_t(d+E_\phi)+\partial_i(v^i+g_\phi^i) &= 0, 
    \label{eq:leading-total-energy}\\
    \partial_t(v^i+g_\phi^i)+c_s^2\partial^i(d+E_\phi) &= \partial_j\Delta^{ij}.
    \label{eq:leading-total-momentum}
\end{align}
These equations describe a forced acoustic evolution of the combined scalar–fluid energy and momentum densities, with the spatial divergence of the stress mismatch acting as an effective driving term. Free acoustic propagation of these combined densities is recovered only if the stress mismatch driving vanishes throughout the subsequent evolution 
\begin{equation}
    \partial_j\Delta^{ij}(t,\mathbf x) = 0, \qquad t\geq t_m,
\end{equation}
This condition is naturally approached at late times, when the scalar remnant has sufficiently relaxed towards the true vacuum and its remaining kinetic, gradient, and potential energies become negligible. The plasma then retains the sound generated during bubble growth and collisions, while the scalar field no longer provides appreciable driving. In practice, instantaneous conversion therefore favors a sufficiently late matching time, potentially requiring the coupled simulation to continue well beyond the collision stage.

By contrast, an earlier instantaneous conversion removes the remaining stress-mismatch driving by hand before scalar relaxation is complete. The discrepancy between the coupled system and the converted free fluid $\delta d=\bar d-(d+E_\phi)$ and $\delta v^i=\bar v^i-(v^i+g_\phi^i)$ follows
\begin{align}
\partial_t\delta d+\partial_i\delta v^i&=0,\notag\\
\partial_t\delta v^i+c_s^2\partial^i\delta d&=-\partial_j\Delta^{ij}.
\label{eq:mismatch-difference}
\end{align}
Thus, even with identical energy and momentum densities at the matching surface $\delta d(t_m)=\delta v^i(t_m)=0$, the two descriptions generally develop different subsequent perturbations. The discrepancy is driven by the remaining spatial variation of the stress mismatch, rather than by any violation of energy-momentum conservation.

\subsection{History dependent acoustic response}

The restriction to late matching can nevertheless be relaxed if the response to the remaining stress mismatch driving can be determined and incorporated into the acoustic initial conditions. As the scalar remnant relaxes, the stress mismatch evolves while the acoustic perturbations continue to propagate. The resulting correction depends on its subsequent history rather than solely on its instantaneous value at $t_m$. 

The acoustic response omitted by instantaneous conversion can be determined from the difference equations in Eq.~\eqref{eq:mismatch-difference}. Working in Fourier space with $\partial_i\to ik_i$, we rewrite each nonzero wavevector $\mathbf{k}=K\mathbf{n}$ with norm $K$ and denote the associated frequency $\omega=c_sK$. The longitudinal projections are denoted by $\delta v_{L,\mathbf k}=\mathbf n\cdot\delta\mathbf v_{\mathbf k}$ and $\Delta_{LL,\mathbf k}=n_i n_j\Delta_{\mathbf k}^{ij}$, with analogous notation for other vectors. 
To diagonalize the coupled density and longitudinal momentum equations, we define the two acoustic branches of the original plasma as 
\begin{equation}
    b_{s,\mathbf k} = c_sd_{\mathbf k}+sv_{L,\mathbf k}
\end{equation}
with $s=\pm1$. Similar definition extends to the converted fluid and the matching discrepancy and is denoted by $\bar b_{s,\mathbf{k}}$ and $\delta b_{s,\mathbf{k}}$, respectively. These combinations separate the longitudinal dynamics into two acoustic branches with opposite propagation phases, whereas the matching discrepancy remains driven by the longitudinal stress mismatch
\begin{equation}
    (\partial_t+is\omega)\delta b_{s,\mathbf k} = -isK\Delta_{LL,\mathbf k}.
\end{equation}
Direct integration from the matching time to a later time $t_h$ solves this equation. 

Together with the free evolution of $\bar b_{s,\mathbf{k}}$, the resolved discrepancy allows the actual acoustic amplitude at later times to be reconstructed as the sum of three contributions:
\begin{equation}
\begin{aligned}
e^{is\omega(t_h-t_m)}b_{s,\mathbf k}(t_h)={}&\bar b_{s,\mathbf k}(t_m)-e^{is\omega(t_h-t_m)}\left[c_sE_{\phi,\mathbf k}(t_h)+sg_{\phi,L,\mathbf k}(t_h)\right]\\
&+isK\int_{t_m}^{t_h}dt\,e^{is\omega(t-t_m)}\Delta_{LL,\mathbf k}(t).
\end{aligned}
\label{eq:finite-history-response}
\end{equation}
The phase factor on the left removes the overall free propagation phase, while that in the integral retains the relative phases of contributions generated at different times. On the right-hand side, the first term represents the acoustic amplitude prescribed by instantaneous conversion at $t_m$ and the second subtracts the contribution of the energy and momentum still carried by the scalar field at $t_h$. The last term captures the history-dependent acoustic response to the remaining stress mismatch, coherently accumulating contributions generated at different stages of scalar relaxation. This response therefore depends on the entire intervening history, rather than solely on the instantaneous mismatch at $t_m$.

At sufficiently late time that the scalar has decayed away, the reconstructed acoustic amplitude motivates the effective acoustic initial condition 
\begin{equation}
\bar b_{s,\mathbf k}^{\rm eff}(t_m)=\bar b_{s,\mathbf k}(t_m)+isK\int_{t_m}^{\infty}dt\,e^{is\omega(t-t_m)}\Delta_{LL,\mathbf k}(t).
\label{eq:effective-acoustic-data}
\end{equation}
This prescription supplements instantaneous conversion with the acoustic response generated during the remaining scalar relaxation, thereby bringing the free propagation approximation into closer agreement with the actual evolution. Concretely, The effective amplitude subsequently evolves freely as
\begin{equation}
    \bar b_{s,\mathbf k}^{\rm eff}(t) = e^{-is\omega(t-t_m)}\bar b_{s,\mathbf k}^{\rm eff}(t_m), 
\end{equation}
and asymptotically reproduces the acoustic amplitude of the original coupled system 
\begin{equation}
    \lim_{t\to\infty}\left[b_{s,\mathbf k}(t)-\bar b_{s,\mathbf k}^{\rm eff}(t)\right]=0.
\end{equation}
This matching property follows directly from the convergence of the stress history integral. Moreover, for a nonnegative scalar potential $U(\varphi_v+\chi)\geq0$, the longitudinal scalar stress satisfies $|T_{LL}^{\phi}|\leq E_\phi$ pointwise. Consequently, the matching error is also bounded for arbitrary time $t\geq t_m$, as
\begin{equation}
\left|\bar b_{s,\mathbf k}^{\rm eff}(t)-\bar b_{s,\mathbf k}(t)\right|\leq(1+c_s^2)K\int_{t_m}^{\infty}dt\int_{\mathbb R^3}d^3x\,E_\phi(t,\mathbf x)
\label{eq:history-response-bound}
\end{equation}
For finite integrated scalar energy, this bound establishes an absolute suppression of the correction at long wavelengths. Its magnitude nevertheless depends on the spatial structure and temporal evolution of the stress mismatch. Moreover, the history correction interferes with the acoustic amplitude prescribed by instantaneous conversion, potentially enhancing or suppressing individual modes. Consequently, a small absolute correction does not necessarily imply a negligible relative change in the acoustic response.

For completeness, we also consider the transverse momentum response. Defining the transverse components
\begin{equation}
    \delta v_{T,\mathbf k}^{i} = \delta v_{\mathbf k}^{i}-n^i\delta v_{L,\mathbf k}, \qquad
    \Delta_{TL,\mathbf k}^{i} = n_j\Delta_{\mathbf k}^{ij}-n^i\Delta_{LL,\mathbf k},
\end{equation}
on can derive their master equation
\begin{equation}
    \partial_t\delta v_{T,\mathbf k}^{i}=-iK\Delta_{TL,\mathbf k}^{i}.
\end{equation}
from Eq.~\eqref{eq:mismatch-difference}. Unlike the longitudinal acoustic modes, the transverse momentum remains constant in the absence of driving at leading order. The corresponding effective initial condition is therefore given by
\begin{equation}
    \bar v_{T,\mathbf k}^{\rm eff,i}(t_m)
    =\bar v_{T,\mathbf k}^{i}(t_m)
    +iK\int_{t_m}^{\infty}dt\,\Delta_{TL,\mathbf k}^{i}(t).
    \label{eq:effective-transverse-data}
\end{equation}
The transverse correction likewise depends on the remaining stress history, but involves no acoustic propagation phase and originates solely from the anisotropic scalar gradient stress.


With the effective initial conditions derived above, free acoustic propagation can reproduce the fluid evolution after scalar field relaxation for late-time observation. We therefore only need to choose a matching time $t_m$ at which the scalar field has entered the relaxation stage.



\section{Computable initial conditions for GW production}
\label{sec:initialcondition}


The effective acoustic initial conditions derived above depend on the subsequent evolution of the scalar field. To determine these conditions, we adopt the harmonic oscillator approximation and calculate the scalar field evolution from its initial configuration and time derivative at $t_m$. In addition, to account for GW production during scalar field relaxation, we derive the corresponding corrections to the initial conditions for the tensor perturbations.

\subsection{Harmonic continuation and computable acoustic data}
\label{sec:harmonic-continuation}

We first recast the effective acoustic initial conditions~\eqref{eq:effective-acoustic-data} in terms of the scalar-fluid energy-momentum transfer. The scalar energy-momentum balance implies that 
\begin{equation}
(\partial_t+is\omega)\left(c_sE_{\phi,\mathbf k}+sg_{\phi,L,\mathbf k}\right)
=isK\Delta_{LL,\mathbf k} -\eta\left[c_s\dot\chi^2-s\dot\chi(\mathbf n\cdot\nabla\chi)\right]_{\mathbf k}.
\label{eq:scalar-acoustic-balance}
\end{equation}
Integrating over the remaining scalar relaxation yields an equivalent expression for the effective acoustic initial conditions: 
\begin{equation}
\bar b_{s,\mathbf k}^{\mathrm{eff}}(t_m)
=b_{s,\mathbf k}(t_m)+\eta\int_{t_m}^{\infty}dt\,e^{is\omega(t-t_m)}
\left[c_s\dot\chi^2-s\dot\chi(\mathbf n\cdot\nabla\chi)\right]_{\mathbf k}(t).
\label{eq:equivalent-acoustic-response}
\end{equation}
This equivalent representation separates the acoustic amplitude already present at $t_m$ from the response generated by subsequent scalar driving. 
In contrast to the stress mismatch formulation, it directly relates the remaining acoustic response to the energy and momentum transferred to the plasma, providing a convenient basis for evaluating the effective initial conditions from the subsequent scalar dynamics.

While the residual fluctuations $\chi$ around the true vacuum $\varphi_v$ becomes sufficiently small, nonlinear self-interactions become negligible and the scalar dynamics reduce to damped harmonic oscillations governed by
\begin{equation}
\ddot\chi_H-\nabla^2\chi_H+\eta\dot\chi_H+M^2\chi_H=0
\label{eq:harmonic-scalar-evolution}
\end{equation}
with $M^2 \equiv U''(\varphi_v)>0$. To ensure continuity at the matching time, the perturbative field $\chi_H$ inherits the same initial condition of $\chi$ at the matching time, namely $\chi_H(t_m)=\chi(t_m)=\chi_m$ and $\dot\chi_H(t_m)=\dot\chi(t_m)=\pi_m$. Being linear with constant coefficients, such an equation admits a closed-form analytical solution. In terms of the elapsed time $\tau \equiv t-t_m$, the spatial Fourier modes decouple and evolve independently as
\begin{equation}
\begin{aligned}
\chi_{H,\mathbf q}(t_m+\tau)
=e^{-\eta\tau/2}\bigg[\chi_{m,\mathbf q}\cos(\Omega_q\tau)+\frac{\pi_{m,\mathbf q}+\eta\chi_{m,\mathbf q}/2}{\Omega_q}\sin(\Omega_q\tau)
\bigg],
\end{aligned}
\label{eq:harmonic-scalar-solution}
\end{equation}
with effective frequency $\Omega_q^2=q^2+M^2-\eta^2/4$. 


Substituting the harmonic solution $\chi_H$ for $\chi$ in Eq.~\eqref{eq:equivalent-acoustic-response}, we obtain an approximation to the history dependent acoustic response, which we refer to as the \textit{harmonic continuation} scheme. The effective acoustic initial conditions are then fully determined by the scalar field and fluid data at $t_m$. Although the quadratic source terms couple different Fourier modes, the time integrals can be evaluated either numerically or analytically in terms of frequency kernels. The resulting effective initial conditions incorporate the acoustic response to the remaining scalar field relaxation, allowing the subsequent acoustic evolution to be described by free propagation without explicitly resolving the short scales associated with the scalar field.

\subsection{Completing the tensor response}
\label{sec:completing-tensor-response}


To extend the above description to GW production, we also need to account for the anisotropic stresses during scalar field relaxation. During this period, the fluid stress in the coupled system differs from that of the freely propagating fluid, and the residual scalar field itself contributes to GW production through its gradient stress. The effects of these stress differences can be incorporated into effective initial conditions for the tensor perturbations at $t_m$.

For each nonzero wavevector $\mathbf k$ and transverse-traceless polarization $\lambda$, the tensor perturbation satisfies
\begin{equation}
\ddot h_{\lambda,\mathbf k}+K^2h_{\lambda,\mathbf k}
=\Pi_{\lambda,\mathbf k},
\qquad
\Pi_{\lambda,\mathbf k}
=16\pi G\,\mathcal P_\lambda^{ij}[T_{ij}]_{\mathbf k},
\label{eq:tensor-wave-equation}
\end{equation}
where $\mathcal P_\lambda^{ij}$ projects onto the normalized polarization tensor. The proposed harmonic continuation scheme indeed enables an explicit analysis of the stress history in accounting the driving and relaxation. Given $\chi_H$ and the corresponding initial condition inherited at $t_m$, the acoustic response $(d_H,\mathbf v_H)$ are fully determined via Eqs.~\eqref{eq:leading-fluid-energy} and \eqref{eq:leading-fluid-momentum}. Meanwhile, the next-to-leading-order scalar back-reaction is also settled down via replacing $U''(\varphi_v+\chi)$ with $M^2$ in Eq.~\eqref{eq:scalar-feedback}. The tensor source is then obtained by the successive reconstructed fluid dynamics. Retaining contributions through ${\cal O}(w_0^{-1})$, it is shown that
\begin{equation}
\Pi_{\lambda,\mathbf k}^{H}=16\pi G\,\mathcal P_\lambda^{ij}\bigg[ \partial_i\chi_H\partial_j\chi_H + \frac{1}{w_0}\bigg(v_{H,i}v_{H,j} + \partial_i\chi_H\partial_j\psi_H + \partial_i\psi_H\partial_j\chi_H\bigg)\bigg]_{\mathbf k}.
\end{equation}
The first term accounts for the direct GW production from the residual scalar gradients, while remaining  terms capture the anisotropic fluid stress and the leading modification of the scalar stress induced by the fluid. 

In contrast, that of the effective free fluid is entirely determined by its momentum
\begin{equation}
    \Pi_{\lambda,\mathbf k}^{H, \mathrm{eff}} = \frac{16\pi G}{w_0}\mathcal P_\lambda^{ij}
\left[ \bar v_{H,i}^{\mathrm{eff}} \bar v_{H,j}^{\mathrm{eff}} \right]_{\mathbf k}.
\end{equation}
This source is also computable with the harmonic continuation scheme. The longitudinal component of $\bar{\mathbf v}_H^{\mathrm{eff}}$ follows from the effective acoustic initial conditions derived in the preceding subsection, while is fixed by the corresponding history dependent matching condition Eq.~\eqref{eq:effective-transverse-data}. Both components can thus be determined from the harmonic continuation, without involving any scalar degrees of freedom. The gap between the two sources is nontrivial, that is characterized by $\delta\Pi^H_{\lambda,\mathbf k}\equiv\Pi_{\lambda,\mathbf k}^{H}-\Pi_{\lambda,\mathbf k}^{H, \mathrm{eff}}$. Its contribution in GW production can be determined, following from the linearity of Eq.~\eqref{eq:tensor-wave-equation}. Similar to matching the acoustic response, such discrepancy can be compensated with effective tensor initial conditions
\begin{equation}
    \begin{aligned}
    h_{\lambda,\mathbf k}^{\mathrm{eff}}(t_m) ={}&h_{\lambda,\mathbf k}(t_m)
    -\frac{1}{K}\int_{t_m}^{\infty}dt\,\sin[K(t-t_m)]\,\delta\Pi_{\lambda,\mathbf k}(t),\\
    \dot h_{\lambda,\mathbf k}^{\mathrm{eff}}(t_m) ={}&\dot h_{\lambda,\mathbf k}(t_m)
    +\int_{t_m}^{\infty}dt\,\cos[K(t-t_m)]\,\delta\Pi_{\lambda,\mathbf k}(t),
    \end{aligned}
    \label{eq:effective-tensor-cauchy-data}
\end{equation}
The original tensor initial conditions account for the GW response accumulated before $t_m$, while the time integrals account for the difference in the GW response between the two descriptions after $t_m$.

With the effective fluid and tensor initial conditions derived above, the late-time evolution of the coupled scalar–fluid system can be finally described by the following equations:
\begin{equation}
    \begin{aligned}
    \partial_t\bar d^{\mathrm{eff}}+\partial_i\bar v^{\mathrm{eff},i}&=0,\\
    \partial_t\bar v_i^{\mathrm{eff}}+c_s^2\partial_i\bar d^{\mathrm{eff}}&=0,\\
    \ddot h_{\lambda,\mathbf k}^{\mathrm{eff}}+K^2h_{\lambda,\mathbf k}^{\mathrm{eff}}
    &=\frac{16\pi G}{w_0}\mathcal P_\lambda^{ij}\left[\bar v_i^{\mathrm{eff}}\bar v_j^{\mathrm{eff}}\right]_{\mathbf k}.
    \end{aligned}
\end{equation}
Within the adopted approximations, the effects of the remaining scalar field relaxation are incorporated into the effective initial conditions at $t_m$. The GW spectrum can then be calculated from the resulting tensor perturbations in the usual manner. In the following section, we will examine the accuracy of the effective acoustic description and its dependence on the matching time using numerical methods.

\section{Numerical simulations and error analysis}
\label{sec:numeric}
In the preceding sections, we developed a perturbative description of the acoustic response to residual scalar-field oscillations and constructed effective initial conditions for the subsequent free propagation of sound waves. However, the spatially inhomogeneous scalar field and fluid configurations at the matching time are determined by the preceding nonlinear dynamics of bubble nucleation, expansion, and collisions, and cannot be obtained from the perturbative framework alone. We therefore extract these configurations from full three-dimensional lattice simulations. As derived in Sec.~\ref{sec:initialcondition}, the subsequent scalar field evolution in the harmonic approximation and its associated acoustic response admit analytical solutions in Fourier space. Their direct application to lattice initial data would require accounting for the lattice dispersion relation and the corresponding discrete spatial operators~\cite{Figueroa:2011ye}. For this reason, we evolve the approximate equations numerically on the same spatial lattice as the full simulations, avoiding additional discrepancies arising from different numerical discretizations. This allows us to test the effective description within a consistent numerical framework.

\subsection{Numerical simulations setup}
To investigate the scalar–fluid dynamics in the post-collision period, we adopt the finite temperature scalar–fluid model of Ref.~\cite{Hindmarsh2015}. The scalar potential is motivated by the high temperature expansion of the finite temperature effective potential in the realistic particle physics models and takes the form
\begin{equation}
V(\phi,T)=
\frac{\gamma}{2}(T^2-T_0^2)\phi^2
-\frac{A}{3}T\phi^3
+\frac{\lambda}{4}\phi^4,
\end{equation}
with $\gamma = 1 /18 $, $A= \sqrt{10}/72$, $\lambda = 10/648$ and $T_0/T_c = 1 / \sqrt{2}$. The temperature dependent quadratic and cubic terms allow the potential to develop a barrier between competing minima, providing a simple effective description of a first-order phase transition without specifying the underlying particle physics model. The corresponding thermodynamic quantities are
\begin{equation}
p=aT^4-V(\phi,T),\qquad
e=3aT^4+V(\phi,T)-T\frac{dV_{T}(\phi,T)}{dT},
\end{equation}
where $a=\pi^2g_*/90$. We solve the coupled scalar–fluid equations introduced in Sec.~\ref{sec:mismatch} with this finite-temperature potential and equation of state, evolving the scalar field \(\phi\), fluid temperature \(T\), and velocity \(\mathbf u\) self-consistently throughout bubble nucleation, collisions, and subsequent relaxation.

The simulations are performed in a three-dimensional periodic box on a Minkowski background. All dimensionful quantities are normalized by the critical temperature $T_c$, with lengths and times measured in units of $T_c^{-1}$, scalar field values and temperatures in units of $T_c$, and energy densities in units of $T_c^4$. We solve the coupled scalar–fluid equations using a staggered-grid scheme with operator splitting, following the numerical approach of Ref.~\cite{Hindmarsh2015}. Spatial derivatives are evaluated using centred finite differences. To accurately capture the propagating fluid shells and the steep gradients arising during bubble collisions, we employ a conservative piecewise-linear upwind scheme with a van Leer limiter, which suppresses spurious numerical oscillations while limiting numerical diffusion~\cite{VANLEER1979101}. The dissipative coupling is treated semi-implicitly, and the local temperature and fluid velocity are reconstructed from the evolved energy and momentum densities.

Alongside the scalar–fluid evolution, we compute the GWs generated by the fluid motion. The corresponding anisotropic stress is constructed from the fluid energy–momentum tensor Eq.~\eqref{eq:emt_fluid}, and its transverse–traceless components are extracted in Fourier space. The two GW polarizations can then be evolved as independent forced oscillators. To maintain consistency with the spatial finite difference scheme, we adopt the lattice dispersion relation

\begin{equation}
\omega_{\mathrm{lat}}^2(\mathbf{k})=
\frac{4}{\Delta x^2}
\sum_{i=1}^{3}
\sin^2\left(\frac{k_i\Delta x}{2}\right).
\end{equation}
The tensor perturbations are evolved throughout bubble growth, collisions, and the subsequent acoustic stage. The GW energy-density spectrum is then computed from the time derivatives of the tensor perturbations by averaging over spherical shells in Fourier space~\cite{Figueroa:2011ye}.

\begin{figure}[t]
    \centering
    \includegraphics[width=\textwidth]{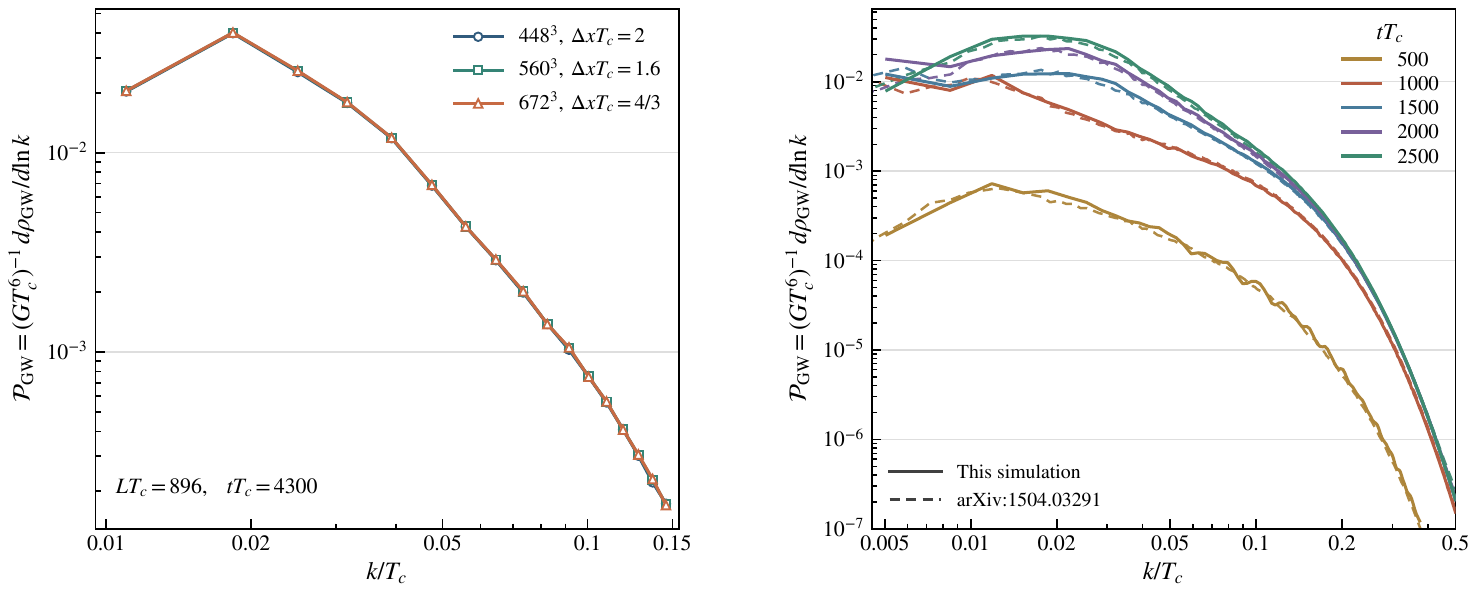}
    \caption{Numerical validation of the GW calculation.
    \textbf{(a)} Resolution convergence of the fluid-sourced
    GW spectrum at $tT_c=4300$, using $448^3$, $560^3$,
    and $672^3$ lattices at fixed volume $LT_c=896$.
    The mean relative deviations from the finest resolution
    are $0.97\%$ and $0.32\%$, respectively.
    \textbf{(b)} Comparison of our total-source GW spectra
    (solid) with those in Fig.~9 of Ref.~\cite{Hindmarsh2015} (dashed) at five different times. Our benchmark uses a $928^3$ lattice with $\Delta xT_c=2$, $\eta/T_c=0.2$, and 59 bubbles.
    }
    \label{fig:gw_validation}
\end{figure}

To validate our numerical implementation, we first examine the convergence of the GW spectrum under spatial lattice refinement. As shown in the left panel of Fig.\ref{fig:gw_validation}, we vary the lattice spacing $\Delta x=L/N$ while keeping the physical box size fixed and using identical nucleation histories. The resulting spectra exhibit excellent agreement over the resolved wavenumber range, with logarithmically weighted mean absolute relative deviations of $0.97\%$ and $0.32\%$ for the $448^3$ and $560^3$ lattices, respectively, relative to the finest $672^3$ lattice. This convergence demonstrates that the GW spectrum is insensitive to further lattice refinement within the range considered.

As an independent validation, we compare our results with the three-dimensional simulations of Ref.~\cite{Hindmarsh2015}. For this benchmark, we include both scalar field and fluid contributions to the GW source, as in the reference calculation. The right panel of Fig.\ref{fig:gw_validation} shows that our results reproduce the time evolution of the GW spectrum, with good agreement around the spectral peak and in the ultraviolet region. Differences are more pronounced at low wavenumbers, where the smaller simulation volume limits the accessible Fourier modes and increases the sensitivity to finite-volume effects. The agreement with previous simulations, together with the numerical convergence demonstrated above, supports the reliability of our subsequent analysis of post-collision dynamics.

\subsection{Numerical tests of perturbative matching and GW predictions}
Having validated the numerical implementation, we now turn to the perturbative matching prescription developed in Sec.~\ref{sec:initialcondition}. We will investigate whether the acoustic response to residual scalar field oscillations can be incorporated into effective initial conditions for subsequent free propagation, and how the resulting GW spectrum depends on the choice of matching time. To examine the prescription under different dynamical conditions, we consider both single bubble and multi bubble configurations.

Both configurations are simulated on a $448^3$ lattice with periodic boundary conditions, a box size $LT_c=896$, a lattice spacing $\Delta xT_c=2$, and a time step $\Delta tT_c=0.1$. We adopt the same scalar potential and fluid equation of state, with an initial temperature $T_n/T_c=0.86$ and a friction coefficient $\eta/T_c=0.1$. In the single bubble configuration, a bubble is initialized at the centre of the simulation volume, providing a relatively simple situation for examining the scalar field relaxation and fluid response. The multi bubble configuration instead contains 13 bubbles nucleated simultaneously at randomly selected positions at $t=0$, allowing us to investigate the evolution of overlapping fluid perturbations.
\begin{figure}[t]
    \centering
    \includegraphics[width=\textwidth]{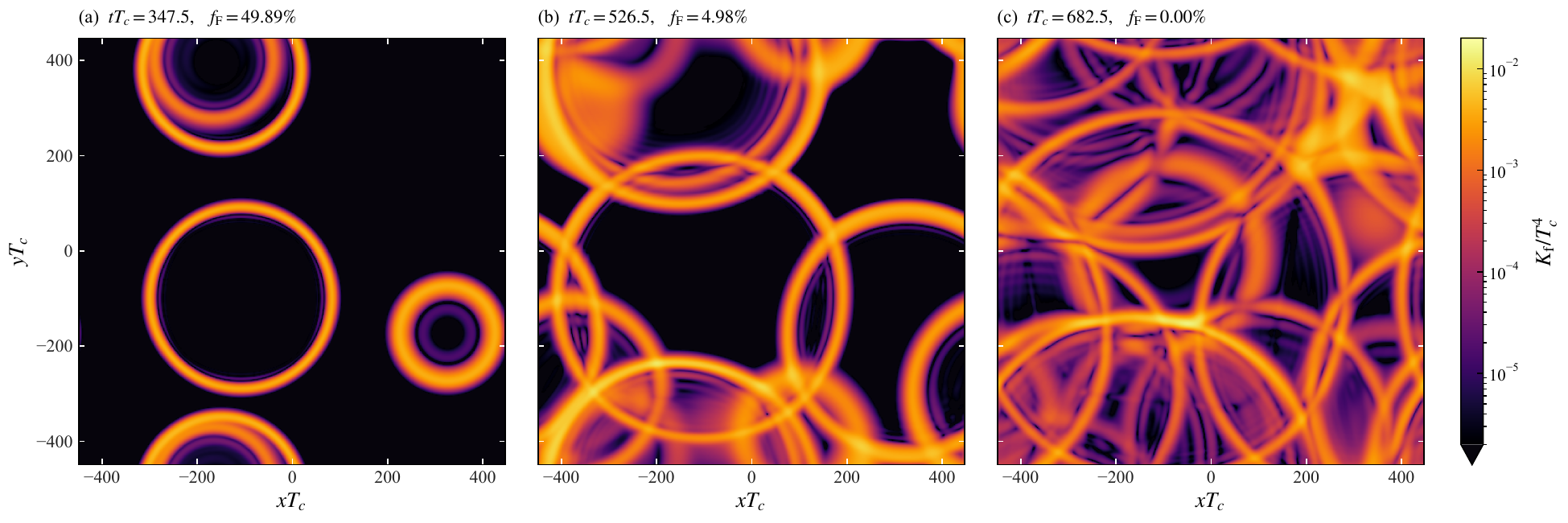}
    \caption{Slices of the fluid kinetic-energy density in the 13-bubble simulation at $tT_c=347.5$, $526.5$, and $682.5$ (left to right). $K_{\rm f}=(\epsilon+p)\gamma^2|\mathbf{u}|^2$, is the fluid kinetic energy density. The annotated $f_{\rm F}$ denotes the instantaneous false vacuum volume fraction of the full three-dimensional box.}
    \label{fig:snapshot}
\end{figure}
To illustrate the resulting dynamics, Fig.~\ref{fig:snapshot} presents representative two-dimensional snapshots of the fluid kinetic energy density in the multi-bubble configuration. The three panels show the same $z=0$ plane at different stages of the evolution, with the corresponding three-dimensional false vacuum volume fraction indicated in each panel.

A central question in applying the perturbative matching prescription is how to choose the matching time. In a realistic phase transition, bubble collisions occur at different times across the simulation volume, making it difficult to identify a unique onset of free acoustic propagation. Matching too early may leave expanding bubble interfaces and large scalar field excursions that are not adequately described by small oscillations about the true vacuum. Matching too late, on the other hand, leaves little residual scalar driving to account for, rendering the matching procedure largely unnecessary. This ambiguity is particularly relevant in multi bubble case, where some regions may already be relaxing while collisions are still taking place elsewhere. Even when all bubbles nucleate simultaneously, their different separations lead to a distribution of collision times.

We therefore examine how the accuracy of the perturbative matching depends on its choice rather than prescribing a particular matching time. For simplicity, we begin with the single bubble configuration, where the onset of collisions can be identified more clearly. As the bubble expands, its wall eventually reaches the boundaries of the periodic simulation volume and collides with its periodic images. The resulting disturbances propagate back through the bubble interior, exciting scalar field oscillations about the true vacuum and continuing to drive the fluid. 

To assess the role of residual scalar driving in the subsequent acoustic evolution, we compare two approximate treatments with the full scalar–fluid simulation. In the first approach, the residual scalar field is removed at $t_m$, and the fluid evolves freely without further driving. This effectively assumes that the sound waves enter the free-propagation stage immediately after bubble collisions. In the second approach, the acoustic response to residual scalar oscillations is incorporated through the perturbative matching prescription developed in Sec.~\ref{sec:initialcondition}. Comparing both predictions with the full evolution allows us to determine how much of the discrepancy introduced by removing the scalar field can be recovered by accounting for its subsequent driving.


At each matching time $t_m$, both approaches are initialized from the same snapshot of the full simulation. We take the volume averaged temperature at $t_m$ as the reference temperature $T_\star$. Since the subsequent acoustic evolution is described in the true vacuum phase, we adopt a homogeneous true vacuum configuration as the background for the matching calculation. Accordingly, the scalar field is replaced by its true-vacuum value $\phi_{\rm t}(T_\star)$, with its time derivative set to zero, while the fluid temperature and velocity are kept unchanged. This projection modifies the fluid enthalpy, so the momentum density is rescaled by the ratio of the new to the original enthalpy to preserve the fluid velocity. Starting from this baseline acoustic state, we consider two different evolutions. In the direct-removal approach, the scalar field and its source terms are discarded, and the fluid evolves freely according to the linear acoustic equations. In the perturbative matching approach, the residual scalar field is described within the damped harmonic approximation, and the resulting acoustic response is incorporated into the baseline state to construct effective acoustic initial conditions at $t_m$. The correction is evaluated by integrating over $[t_m,t_h]$, with a common upper integration limit $t_h$ for all matching times within each configuration. The subsequent fluid evolution for $t\geq t_h$ is then described by free acoustic propagation from these effective initial conditions. In both approaches, the tensor perturbations and their time derivatives at $t_m$ are taken directly from the full simulation. The subsequent GW evolution is sourced solely by the fluid anisotropic stress, so that the effects of residual scalar field oscillations enter the GW calculation only through the corrected acoustic initial conditions in the perturbative matching approach. We compare the resulting GW spectra with those obtained from the full simulation at the same evaluation times and quantify their relative deviations.

\begin{figure}[t]
    \centering
    \includegraphics[width=\textwidth]{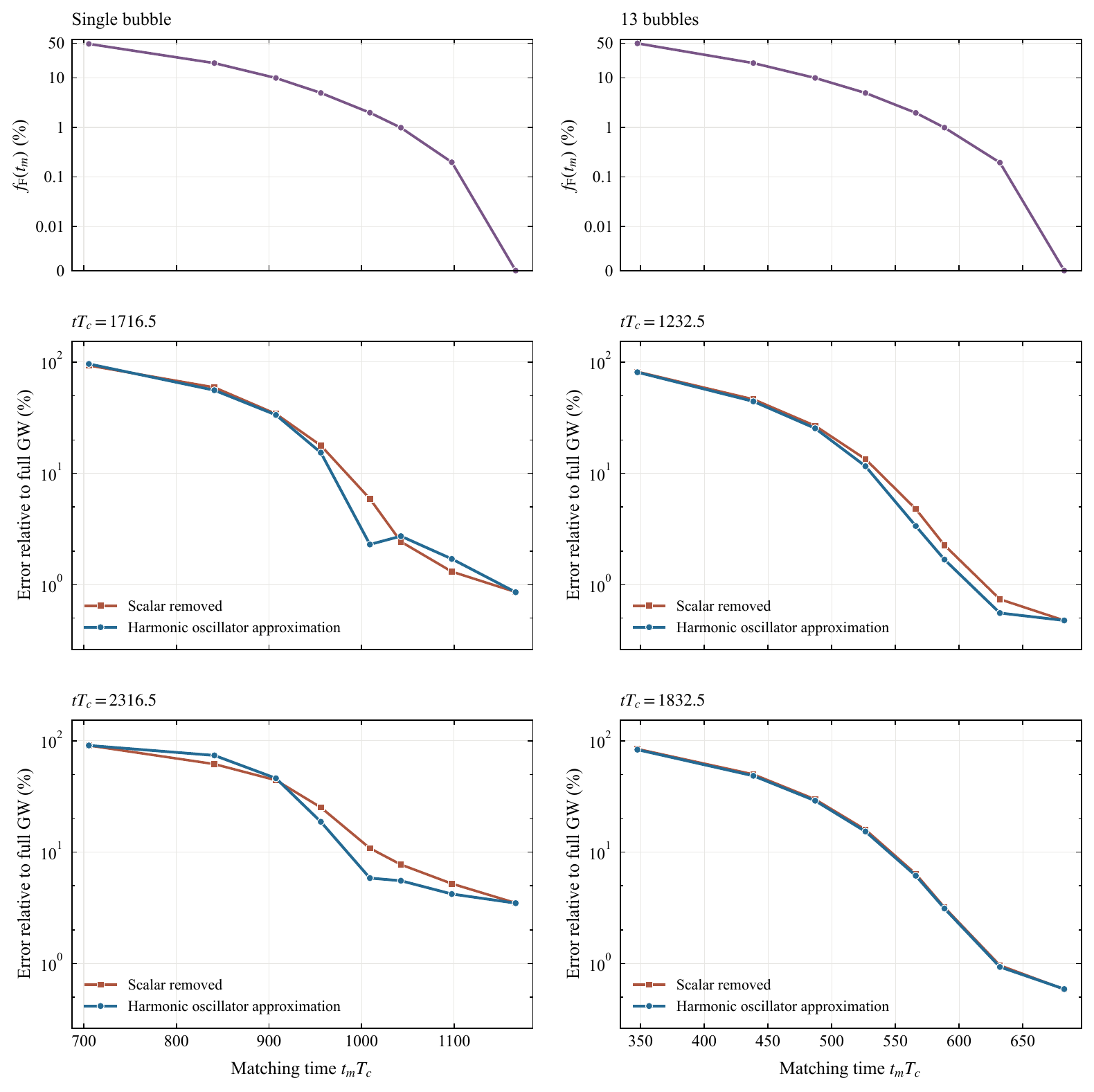}
    \caption{False vacuum volume fraction (top) and GW spectral errors relative to the full simulation (middle and bottom) as functions of the matching time. Results for the single bubble (left) and multi bubble (right) configurations compare direct scalar removal with the harmonic oscillator approximation at the indicated readout times.}
    \label{fig:matching_error}
\end{figure}

Fig.~\ref{fig:matching_error} shows the dependence of the GW spectral errors on the matching time for the single bubble (left) and multi bubble (right) configurations. The upper panels track the false vacuum volume fraction, while the middle and lower panels compare the two approximations at different evaluation times. For the single bubble configuration, both approaches exhibit large errors at early matching times, when bubble collisions and substantial scalar field excursions are still present. As the transition proceeds, the errors decrease rapidly, indicating that the subsequent evolution becomes increasingly well described by the fluid alone. The advantage of perturbative matching becomes particularly apparent when the false vacuum volume fraction falls to a few percent. At $t_mT_c\simeq1010$, where approximately $2\%$ of the volume remains in the false vacuum, perturbative matching reduces the GW spectral error from about $6\%$ to $2\%$ at $tT_c=1716.5$, and from $11\%$ to $6\%$ at $tT_c=2316.5$. This improvement suggests that the damped harmonic approximation captures a significant part of the residual scalar driving once most of the field has approached the true vacuum.

As the matching time is delayed further, the errors of both approaches generally decrease, although their behaviour differs between the two evaluation times. At $tT_c=1716.5$, both errors eventually fall below $1\%$, indicating that the effects neglected by the approximate evolution have only a small influence on the GW spectrum over this period. At the later evaluation time $tT_c=2316.5$, however, a discrepancy of approximately $3$--$4\%$ persists even after the false vacuum volume fraction has vanished. This difference suggests that corrections neglected in the approximate description, potentially including nonlinearities in the scalar potential and the backreaction of the fluid on the scalar field, may become more important over longer evolution times. Therefore their possible cumulative effects deserve further consideration in predictions of late-time GW production.

\begin{figure}[t]
    \centering
    \includegraphics[width=\textwidth]{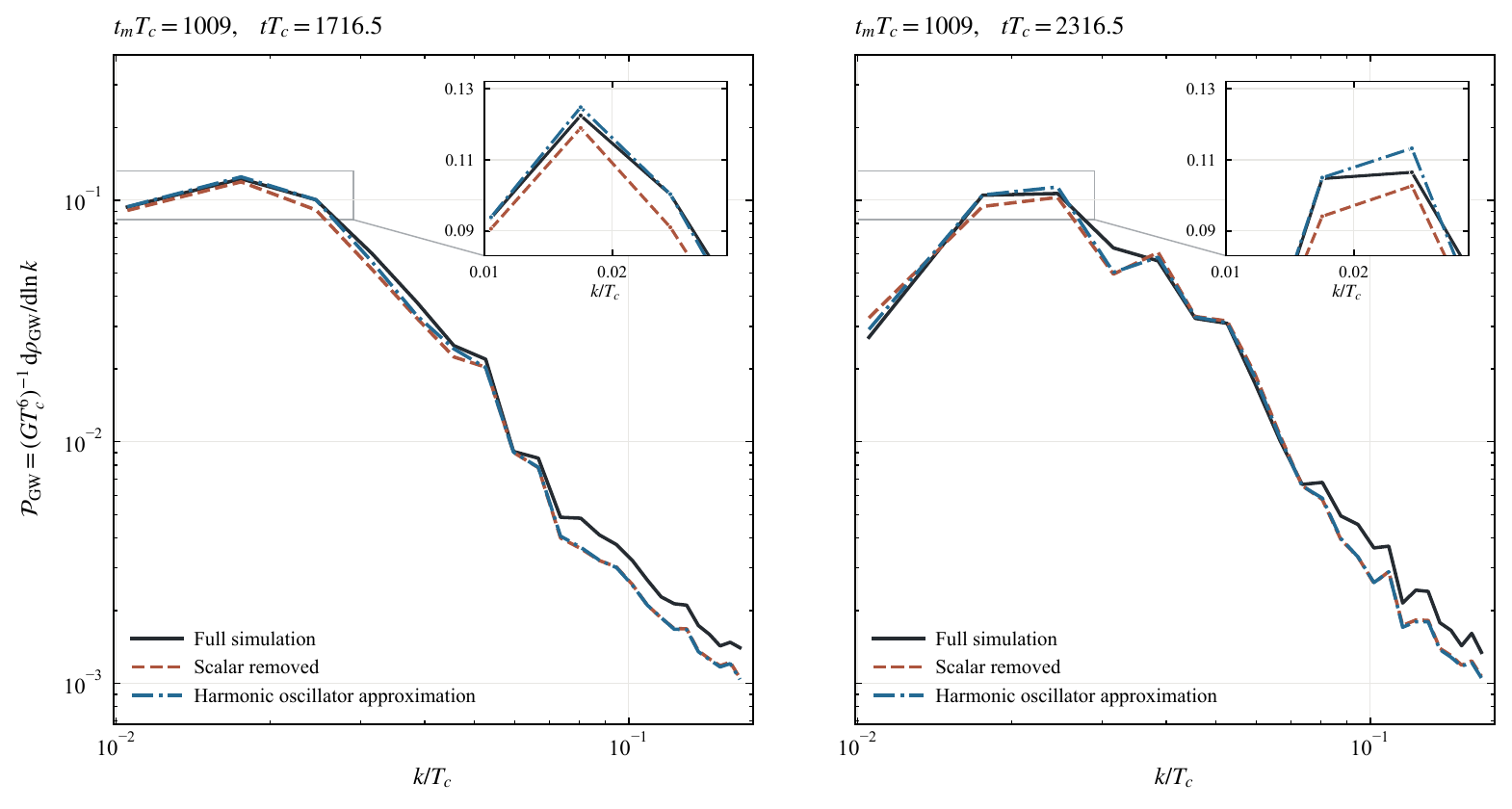}
    \caption{GW spectra for the single-bubble configuration matched at $t_mT_c=1009$, evaluated at $tT_c=1716.5$ (left) and $2316.5$ (right). The full simulation is compared with direct scalar removal and the harmonic oscillator approximation. Insets show the peak region.}
    \label{fig:single_gw}
\end{figure}

To identify the spectral origin of the differences between the two approaches, we further examine the GW spectra at two evaluation times. We choose $t_mT_c=1009$, when approximately $2\%$ of the volume remains in the false vacuum, and compare the predictions with the full simulation at $tT_c=1716.5$ and $2316.5$, as shown in Fig.~\ref{fig:single_gw}. The perturbative matching reproduces the spectral peak more accurately than direct removal, particularly at the later evaluation time. This improvement indicates that the damped harmonic approximation captures an important part of the residual scalar driving that contributes to GW production on the dominant acoustic scales. At higher wavenumbers, however, both approximations give nearly identical spectra and systematically underestimate the full simulation, indicating that the dynamics responsible for part of the small-scale GW production are not captured by the present effective description.

The single bubble results show that the harmonic approximation can recover part of the residual scalar driving, particularly its contribution to the GW spectrum near the acoustic peak. It is therefore interesting to examine whether this improvement persists in a multi bubble configuration. As shown in the right panels of Fig.~\ref{fig:matching_error}, the GW spectral errors again decrease as the transition proceeds. When the false vacuum volume fraction reaches approximately $1\%$, at $t_mT_c\simeq590$, the matched spectra agree with the full simulation within about $2\%$ and $3\%$ at $tT_c=1232.5$ and $1832.5$, respectively. Unlike the single bubble case, however, the improvement over direct removal is much less pronounced, particularly at the later evaluation time.

Unlike in the single bubble case, bubble collisions in the multi bubble configuration occur at different times, leaving the scalar field at different stages of relaxation at a given matching time. Even when the false vacuum volume fraction becomes small, the scalar field may still exhibit large deviations from the true vacuum in some regions. The validity of the harmonic oscillator approximation therefore depends on the local relaxation of the scalar field rather than simply on the overall progress of the phase transition.

The residual discrepancies in the GW spectra, particularly in the ultraviolet region and at later evaluation times, motivate us to consider nonlinear corrections beyond the harmonic approximation. As a possible extension, we retain the quadratic force term while neglecting the cubic force term arising from the quartic interaction in the potential. The residual scalar equation then takes the form
\begin{equation}
\mathcal D_\phi\chi=-g_2\chi^2,
\qquad
\mathcal D_\phi=
\partial_\tau^2+\eta\partial_\tau-\nabla^2+M^2.
\end{equation}
Introducing a formal expansion parameter $\epsilon$, we write $\chi=\sum_{n\geq1}\epsilon^n\chi_n$. Substituting this expansion into the equation of motion and collecting terms order by order gives
\begin{equation}
\mathcal D_\phi\chi_1=0,\qquad
\mathcal D_\phi\chi_n
=-g_2\sum_{j=1}^{n-1}\chi_j\chi_{n-j},
\qquad n\geq2.
\end{equation}
The leading-order solution carries the prescribed initial data, while all higher-order corrections have vanishing initial data. Each successive correction is therefore governed by the same linear operator, with a source constructed from lower-order solutions. In particular, the first nonlinear correction satisfies $\mathcal D_\phi\chi_2=-g_2\chi_1^2$. This hierarchy provides a systematic extension beyond the harmonic approximation. Its numerical implementation requires solving additional three-dimensional equations for the higher-order scalar field perturbations, while a consistent treatment of fluid backreaction requires extending the perturbative expansion to the coupled scalar–fluid system. Although the present analysis is restricted to leading order, the perturbative framework establishes a theoretical basis for systematically improving the description of the transition from bubble collisions to free acoustic propagation.


\section{Conclusions}
\label{sec:conclusions}
In this work, we have investigated the effects of residual scalar field oscillations on the transition from bubble collisions to free acoustic propagation in cosmological first-order phase transitions. Our main conclusions are as follows
\begin{itemize}
    \item We establish a perturbative framework for describing the acoustic response to residual scalar field oscillations after bubble collisions. Within the harmonic approximation, the effects of the subsequent energy and momentum transfer from the scalar field to the fluid can be incorporated into effective initial conditions for free acoustic propagation.
    \item Using three-dimensional lattice simulations, we demonstrate the validity of this framework in the single bubble configuration, where the effective initial conditions reproduce the subsequent GW spectrum with errors of a few percent near the end of the phase transition. The improvement is less pronounced in the multi bubble case, where collisions occurring at different times complicate the relaxation of the scalar field and the applicability of the harmonic approximation.
\end{itemize}
Our results provide a theoretical connection between bubble collision dynamics and the subsequent acoustic stage, contributing to a more complete understanding of cosmological first-order phase transition dynamics and the associated GW production.

\acknowledgments
This work was supported by the National Natural Science Foundation of China (Grant Nos.12335005 and 12605177),  the National Natural Science Foundation of Henan province (Grant Nos. 27AQ140008 and 262300421233).


\bibliographystyle{JHEP}
\bibliography{references}

\end{document}